\documentclass[aps,nofootinbib,twocolumn,floatfix,superscriptaddress,secnumarabic]{revtex4-1}
\usepackage{amsfonts,amssymb,epsfig,verbatim,mathbbol,amstext,amsmath,psfrag}
\usepackage{graphicx}
\usepackage{t1enc}
\usepackage{amsmath}
\usepackage{subfigure}
\usepackage{dsfont}
\usepackage{slashed}

\usepackage{colordvi}
\usepackage{xcolor}
\usepackage{color}

\newif\ifSMver
\SMverfalse

\ifSMver

\newcommand{\appendixsection}{%
\setcounter{equation}{0}
\setcounter{section}{0}
\newpage
\renewcommand{\theequation}{S\arabic{equation}}
\section*{Supplemental Material}}
\else

\newcommand{\appendixsection}{\appendix}
\fi

\newcommand{\be}{\begin{equation}}
\newcommand{\ee}{\end{equation}}

\newcommand{\Z}{\mathcal{Z}}

\newcommand{\expv}[1]{\left \langle #1 \right \rangle}
\newcommand{\tr}{\textmd{tr}\,}

\newcommand{\Dp}{\slashed{D}(\mu_I)}
\newcommand{\light}{{ud}}
\newcommand{\Nxi}{N}
\newcommand{\deltaNxi}{\delta^{\Nxi}}
\newcommand{\dNn}{\deltaNxi}

\usepackage[    bookmarks,
                 bookmarksopen = true,
                 bookmarksnumbered = true,
                 breaklinks = true,
                 linktocpage,
                 colorlinks = true,
                 linkcolor = blue,
                 urlcolor  = blue,
                 citecolor = blue,
                 anchorcolor = green,
                 hyperindex = true,
                 hyperfigures]
		{hyperref}        
        
\newif\ifmakewordcount
\makewordcountfalse

\ifmakewordcount

\usepackage{xesearch}
\newcounter{totalwords}
\AtBeginDocument{
    \setcounter{totalwords}{0}
    \SearchList!{wordcount}{\stepcounter{totalwords}}
        {a?,b?,c?,d?,e?,f?,g?,h?,i?,j?,k?,l?,m?,n?,o?,p?,q?,r?,s?,t?,u?,v?,w?,x?,y?,z?}
    \UndoBoundary{'’-}
    \MakeBoundary{„“‚‘–}
    \SearchOrder{p;}
}
\AtEndDocument{
    \StopSearching
    \newpage \thispagestyle{empty}
    {\color{blue}{\arabic{totalwords} words in total}}
}
\newcounter{totwords}
\newcounter{words}
\newenvironment{counted}{%
   \setcounter{words}{0}
   \SearchList!{wordcount2}{\stepcounter{words}}
     {a?,b?,c?,d?,e?,f?,g?,h?,i?,j?,k?,l?,m?,
     n?,o?,p?,q?,r?,s?,t?,u?,v?,w?,x?,y?,z?}
   \UndoBoundary{'}
   \SearchOrder{p;}}{%
   \StopSearching
   \addtocounter{totwords}{\value{words}}
   }
   \newcommand{\wc}{{\color{brown}{+\thewords=\thetotwords} words}}
\else
\newenvironment{counted}{}{}
\newcommand{\wc}{}
\fi

\begin{document}
\title{Reliability of Taylor expansions in QCD}

\author{B.~B.~Brandt}
\affiliation{Institute for Theoretical Physics, Goethe Universit\"at Frankfurt, D-60438 Frankfurt am Main, Germany}
\author{G.~Endr\H{o}di}
\affiliation{Institute for Theoretical Physics, Goethe Universit\"at Frankfurt, D-60438 Frankfurt am Main, Germany}

\begin{abstract}
We investigate the reliability of the Taylor expansion method in QCD with 
isospin chemical potentials using lattice simulations. By comparing the 
expansion of the number density to direct results, the 
range of validity of the leading- and next-to-leading order expansions
is determined. 
We also elaborate on the convergence properties of the Taylor series 
by comparing the leading estimate for the radius of convergence 
to the position of the nearest singularity, i.e.\ the 
onset of pion condensation. 
Our results provide a handle for quantifying the uncertainties of
Taylor expansions in baryon chemical potentials. 
\end{abstract}

\pacs{some pacs} 

\maketitle

\section{Introduction}

\begin{counted}

The thermodynamic properties of QCD at finite temperature and density are in the focus of current research in theoretical
and experimental physics and are of fundamental relevance for the structure of compact stars and the
evolution of the early universe.
On the theoretical side, simulations of lattice QCD are the preferred non-perturbative tool to investigate
the properties of QCD at strong coupling. Monte-Carlo simulations of lattice QCD, however, are hindered for nonzero 
baryon density $n_B$ due to the well known complex
action problem (for recent reviews see~\cite{deForcrand:2010ys,Aarts:2013lcm}).
Despite a number of proposals for methods to potentially overcome this problem and
to enable direct simulations in this regime (for reviews see Ref.~\cite{Gupta:2011ma,Gattringer:2014nxa,Ding:2017giu}, for instance) there is currently no
method which can provide reliable results in the interesting region with $T\lesssim T_{pc}$ at physical quark masses.
Here $T_{pc}$ is the crossover, or pseudo-critical, temperature associated with effective chiral symmetry restoration.

Nonzero-density QCD is studied most conveniently in the grand canonical ensemble, where the baryon density
is traded for the associated chemical potential $\mu_B$.
One approach to circumvent the complex action problem is based
on a Taylor expansion of observables in powers $\mu_B$ at zero chemical potential.
Following the pioneering works~\cite{Gottlieb:1988cq,Gavai:2001fr,Allton:2002zi},
today the Taylor expansion method is the most established approach~\cite{Bazavov:2017dus,Borsanyi:2018grb} 
to investigate regions of the finite-density phase diagram relevant for heavy-ion phenomenology.
However, since the expansion can only be carried out to a
finite order $n$ (currently typically $n\le 8$), the region of reliability 
of the series is {\it a priori} unknown, leaving systematic uncertainties due to higher orders 
difficult to estimate.

Another piece of information encoded in the Taylor expansion coefficients is the potential existence
of a singularity in the complex $\mu_B$-plane -- for example a phase 
transition at real critical chemical potential $\mu_{B,c}$. Due to the
non-analyticity at $\mu_{B,c}$, the phase transition cannot be described by a series expansion in one of
the adjacent phases. In turn, this shows up as a finite radius of convergence for the series expansion
of observables~\cite{Gavai:2004sd}.
This method has been applied extensively in QCD to probe the presence of a possible second order critical
endpoint in the $\mu_B-T$
plane, see, e.g., Refs.~\cite{Allton:2005gk,Gavai:2008zr,DElia:2016jqh,Datta:2016ukp}.

A similar expansion can also be applied in the case of QCD at finite isospin chemical potential
$\mu_I$~\cite{Allton:2005gk,Borsanyi:2011sw}, which is also realized in the aforementioned physical systems.
One advantage of QCD with a pure isospin chemical potential
(i.e.\ $\mu_I\neq0$ but $\mu_B=0$) is that the
complex action problem is absent and the theory can be simulated with standard Monte-Carlo methods~\cite{Kogut:2002tm}.
Consequently, QCD at pure
isospin chemical potential can serve as a realistic test system to investigate the range of applicability of the Taylor
expansion method.

After the initial lattice studies of QCD at $\mu_I>0$ and $\mu_B=0$ at finite lattice spacings and unphysical pion
masses~\cite{Kogut:2002tm,Kogut:2002zg,Kogut:2004zg,deForcrand:2007uz,Detmold:2012wc,Cea:2012ev,Endrodi:2014lja},
we have recently determined its continuum phase diagram~\cite{Brandt:2017oyy}. 
It has a rich structure: besides the chirally broken and restored regions at low chemical 
potential, it exhibits a Bose-Einstein condensed (BEC) phase of charged 
pions beyond a critical chemical potential $\mu_{I,c}(T)$. 
According to our findings, the boundary of the BEC phase is at $\mu_{I,c}\approx m_\pi/2$ for
temperatures up to about $150\textmd{ MeV}$. This is followed by a pronounced turn and a saturation 
at around $T\approx 160 \textmd{ MeV}$ for chemical potentials $\mu_I\leq120$~MeV. The appearance
of the BEC phase is accompanied by the spontaneous breaking of the residual $\mathrm{U}_{\tau_3}\!(1)$
symmetry, remaining from the chiral $\mathrm{SU}_V(2)$ symmetry group at finite $\mu_I$.
Consequently, the phase transition to the BEC phase is expected to be of second order in the
$\mathrm{O}(2)$ universality class~\cite{Son:2000xc}, which is consistent with the finite
volume-dependence and the critical scaling of the lattice results~\cite{Brandt:2017oyy}.

In this letter, we extend the simulations of~\cite{Brandt:2017oyy} 
to test the performance of Taylor expansion in $\mu_I$.
In particular, we investigate the applicability of the Taylor expansion method 
for a broad range of temperatures and study its
capability to determine the BEC phase boundary via estimates of the 
radius of convergence.

\end{counted}\wc

\section{Lattice setup}
\label{sec:setup}

\begin{counted}

We employ the tree-level Symanzik improved gluon action and 2+1 flavors of rooted staggered quarks with two-levels
of stout smearing at physical quark masses, following the line of constant physics from~\cite{Borsanyi:2010cj}.
The continuum limit is approached using lattice ensembles with $N_t=6,8,10$ and $12$, 
corresponding to lattice spacings of $a=0.20,0.15,0.12$ and $0.10\:\textmd{fm}$ around the zero-density 
crossover temperature $T_{pc}(\mu_I=0)$. To enable the observation
of the spontaneous breaking of the $\mathrm{U}_{\tau_3}\!(1)$ symmetry in finite volumes and to regulate
the theory in the infrared, we introduce
a pionic source $\lambda$ in the fermion matrix for the light quark masses.
This source term leads to an unphysical explicit breaking of the $\mathrm{U}_{\tau_3}\!(1)$
symmetry and physical results are obtained in the limit $\lambda\to0$. For a more detailed discussion
see~\cite{Brandt:2017oyy}.

Our main observable is the isospin density
\be
\label{eq:nI-def}
\expv{n_I} = \frac{T}{V}\frac{\partial \log\Z}{\partial \mu_I}\,,
\ee
which is free of ultraviolet divergences and, thus, does not require renormalization.
Its computation in terms of lattice operators is discussed in detail in App.~\ref{app:nI}. The most difficult task
for a reliable computation of $\expv{n_I}$ is the extrapolation in $\lambda$. 
Similarly to our experience with other observables~\cite{Brandt:2017oyy}, the $\lambda$-dependence
of $\expv{n_I}$ is very pronounced, so that a
naive extrapolation cannot be performed in a controlled manner. In Ref.~\cite{Brandt:2017oyy}
we introduced an improvement program
for the $\lambda$-extrapolations, using the singular values of the massive Dirac operator.
Similar improvements can be applied to $\expv{n_I}$ as well, and we discuss the details in App.~\ref{app:nI}.
The dependence on $\lambda$ is reduced substantially, allowing for fits of the data to a constant or
a linear function in $\lambda^2$ (note that $\expv{n_I}$ is an even function of $\lambda$
due to $\mathrm{U}_{\tau_3}(1)$ symmetry).

The Taylor expansion for the isospin density with respect to $\mu_I/T$ is given by
\be
\label{eq:nI-taylor}
\frac{\left\langle n_I \right\rangle
}{T^3} = c_2 \Big(\frac{\mu_I}{T}\Big) +
\frac{c_4}{6} \Big(\frac{\mu_I}{T}\Big)^3 +\ldots\,,
\ee
where $c_2$ and $c_4$ are the associated Taylor coefficients. In the following 
we will consider the leading order $\expv{n_I}^{\rm LO}$ (including $c_2$) 
and the next-to-leading order $\expv{n_I}^{\rm NLO}$ (including $c_2$ and $c_4$) series. For our
action and temporal extents $N_t$, the Taylor expansion coefficients have been computed in
Ref.~\cite{Borsanyi:2011sw}, albeit at different temperatures and, in some cases, on slightly
different volumes. To arrive at the temperatures used in our study, we have performed a cubic spline
interpolation of the associated results. In addition, we found the volume dependence of the Taylor
coefficients to be sufficiently small, so that the effects due to the slight differences in volume
are negligible. The details of the interpolations and the study of volume effects are provided in
App.~\ref{app:ctaylor}.

\end{counted}\wc

\section{Testing the Taylor expansion against direct results}
\label{sec:comp}

\begin{figure}[t]
 \centering
 \includegraphics[width=8cm]{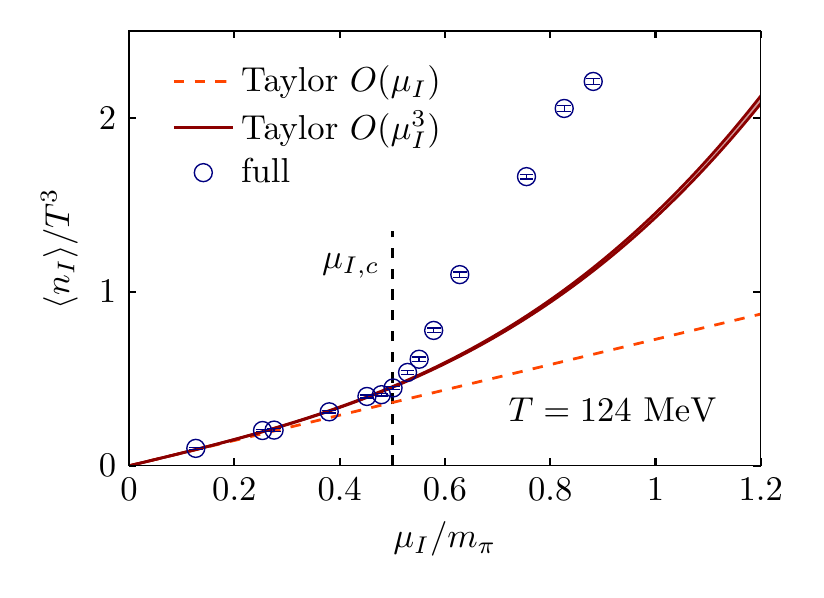}\quad
 \includegraphics[width=8cm]{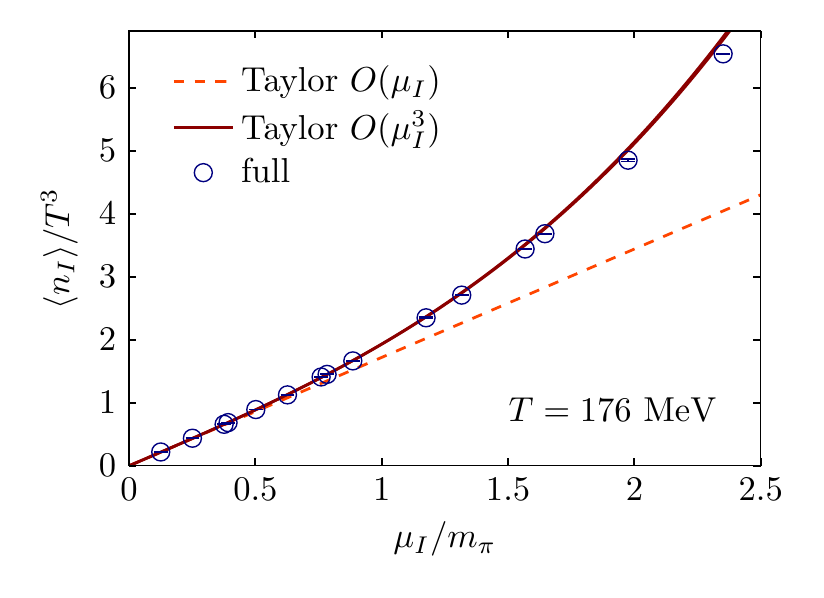}
 \caption{\label{fig:nt6-temps}
 Results for $\expv{n_I}$ on $N_t=6$ lattices from direct simulations (blue points)
 in comparison to LO (orange dashed line) and NLO (red solid line) Taylor expansions
 for $T=124 \textmd{ MeV}$ (top panel) and for $T=176$~MeV (bottom panel).
 }
\end{figure}

\begin{counted}

To perform a detailed comparison between the direct results for $\expv{n_I}$ and the Taylor
expansion in a wide range of $\mu_I$, we have extended our existing results~\cite{Brandt:2017oyy} 
with new data up to $\mu_I\lesssim325$~MeV. This value is still sufficiently far away
from the saturation region for $N_t\geq8$, ensuring that lattice artifacts remain under control
(for all the results shown below, the chemical potential in lattice units satisfies $\mu_Ia < 0.3$).
The comparison between the direct data for $\expv{n_I}$ and the results from
the Taylor expansion is shown in Fig.~\ref{fig:nt6-temps} for $N_t=6$. 
For the lower temperature,
$T=124$~MeV, the data reaches the BEC phase boundary at $\mu_{I,c}\approx m_\pi/2$.
Up to this point the data shows remarkable agreement with both the LO and the NLO 
Taylor expansion. Slight differences between the data and the LO expansion become
apparent close to $\mu_{I,c}$ (see also Fig.~\ref{fig:tcoeff} in App.~\ref{app:ctaylor} at $T=113$~MeV).
The lattice data starts to deviate from the
Taylor expansion curve for $\mu_I>\mu_{I,c}$. This is certainly expected, since the Taylor expansion
cannot capture the change of dynamics at the transition.
In contrast, the results for $T=176$~MeV are above the BEC phase boundary. In this regime the agreement
with Taylor expansion persists up to larger values of $\mu_I$ and starting at around
$\mu_I/m_\pi\approx 0.6$ one can clearly see that the data favors NLO Taylor expansion over the LO
expansion. Furthermore, Taylor expansion at NLO fails to describe $\expv{n_I}$ within the current
uncertainties at around $\mu_I/m_\pi\approx 1.6$, where higher orders become
important for this temperature.

Comparing the behavior of the data points between the two temperatures reveals another characteristic
of the Taylor expansion. For low temperatures, the NLO expansion tends to underestimate
the results for $\expv{n_I}$, while it overestimates them for higher temperatures. 
Due to continuity there will be a region,
where the agreement between the NLO expansion and the full result is almost perfect.
Outside the BEC phase, where the expansion converges to the correct 
result, this region is related to the suppression of higher order terms, most dominantly of
$c_6(T)$, which is indeed expected to cross zero as $T$ increases.
Inside the pion condensation phase this agreement is merely accidental and does not reveal
any information about the interior of the BEC region.

\end{counted}\wc

\begin{figure}[t]
 \centering
 \includegraphics[width=8cm]{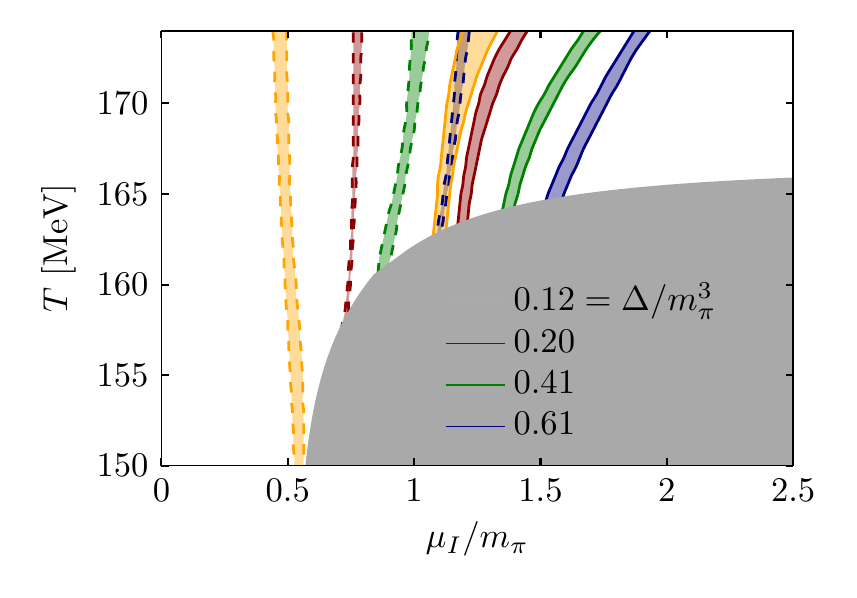}
 \caption{\label{fig:tcontour-nt8}
 Contours of constant $\Delta^{\rm LO}$ (dashed bands) and $\Delta^{\rm NLO}$
 (solid bands) for $N_t=8$. The shaded gray area indicates the BEC phase.
 }
\end{figure}

\begin{counted}

To quantify the regions in parameter space where Taylor expansion at a given order (LO or NLO) starts to
become unreliable, we look at curves in parameter space with constant difference
\be
\label{eq:taylor-dif}
\Delta^{\rm LO/NLO} = \big\vert \expv{n_I} - \expv{n_I}^{\rm LO/NLO}
\big\vert
\ee
between the full results and the Taylor expansion.
In the following we focus on the high temperature
region, $T\gtrsim150$~MeV, to be able to draw conclusions about the applicability
range of the Taylor method in the absence of the BEC phase transition. 

The contour lines are determined using a
two-dimensional spline fit to $\Delta$, where the nodepoints have been generated via a 
Monte-Carlo analysis (for a description of our fit strategy, see Ref.~\cite{Brandt:2016zdy}).
In the spline fit we include the constraint that $\Delta=0$ for $\mu_I=0$. Note that we expect
a rapid change of the data for $\Delta$ at the BEC phase boundary in the thermodynamic limit.
For our finite volumes, for $T\gtrsim150$~MeV the behavior is more regular and can be captured by a spline
interpolation.\footnote{This is similar to the behavior of the chiral condensate reported
in Ref.~\cite{Brandt:2017oyy}.}

Our $N_t=8$ results for the contour lines are shown in Fig.~\ref{fig:tcontour-nt8} for various values of $\Delta$. The figure also includes the BEC phase boundary, which we extended to higher values 
of $\mu_I$ compared to Ref.~\cite{Brandt:2017oyy}, see App.~\ref{app:pcbound} for details.
Most of the contour lines have positive slopes, indicating the general tendency
that the expansion performs better and better as the temperature increases.
This is partly due to the fact that the actual dimensionless expansion parameter is $\mu_I/T$,
cf.\ Eq.~(\ref{eq:nI-taylor}) -- however, the contour lines differ from 
the simple $\mu/T=\textmd{const}.$ lines considerably (see below). 
The exception is the contour line with $\Delta^{\rm LO}/m_\pi^3=0.12$, which 
is roughly insensitive to the temperature.
In addition, the results clearly reflect that the NLO expansion has a broader reliability
range than the LO one, with contour lines shifted to considerably higher
values of $\mu_I$.

\end{counted}\wc

\begin{figure}[t]
 \centering
  \vspace*{-.2cm}
 \includegraphics[width=8cm]{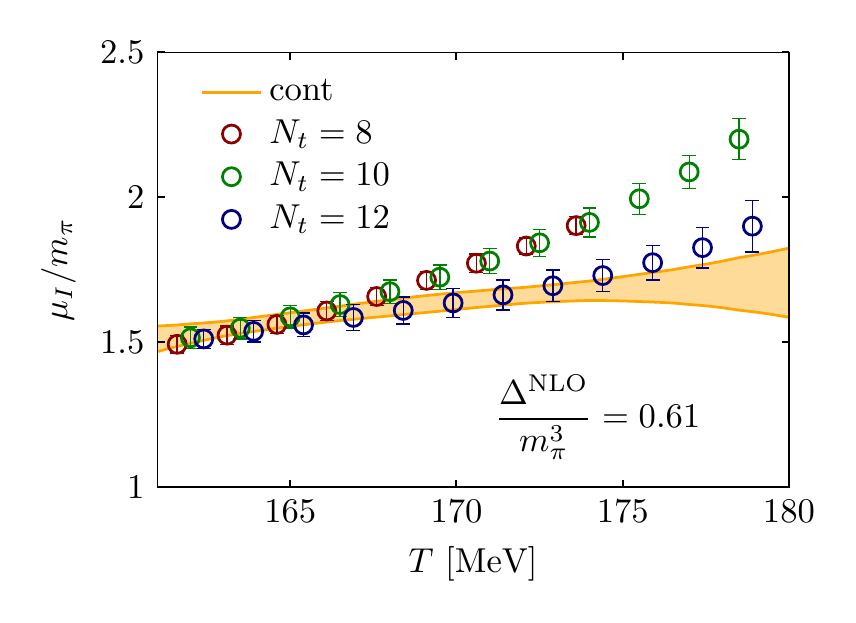}
 \caption{\label{fig:tconti}
 Continuum extrapolation for the contour line $\Delta^{\rm NLO}/m_\pi^3=0.61$. The yellow curve
 corresponds to the continuum extrapolation and the points show (every third of) the results
 from the individual lattices that were included in the fit. We have slightly shifted the data
 horizontally to enhance visibility.
 }
\end{figure}

\begin{figure}[t]
 \centering
 \vspace*{-.05cm}
 \includegraphics[width=7.7cm]{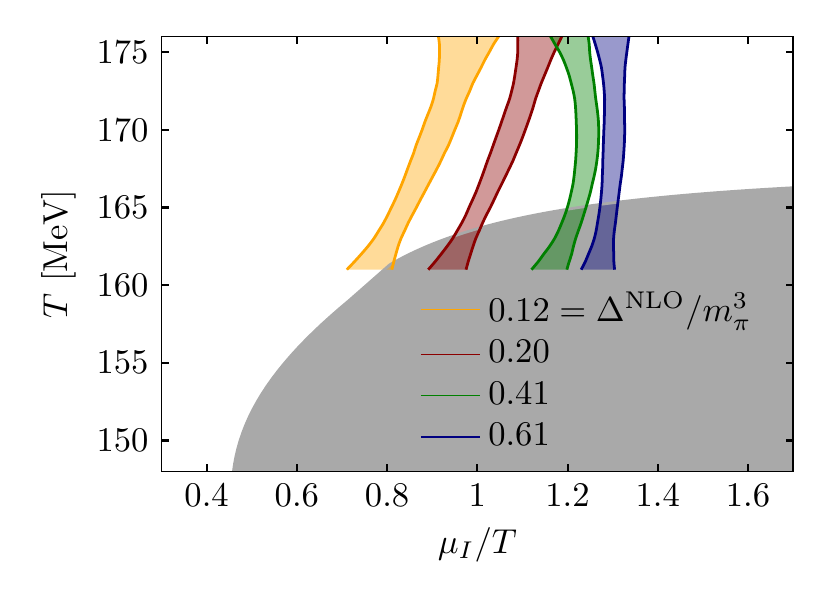}
 \caption{\label{fig:tcontour-conti}
 Continuum results for the contour lines of $\Delta^{\rm NLO}$. The gray shaded area 
 indicates the BEC phase. Note that the curves stop at $T=161$~MeV, since only data
 above this temperature enters the continuum fit.
 }
\end{figure}

\begin{counted}

Eventually we aim at investigating the range of applicability of the Taylor expansion in the continuum. To
this end we perform a continuum extrapolation of the contour lines of $\Delta$, using a parameterization
in terms of polynomials in $(T-T_0)$ with lattice spacing dependent coefficients, setting $T_0=140$ MeV.
In the continuum extrapolation we focus on $\Delta^{\rm NLO}$, for which the contours are well described
by a second order polynomial for $T\geq161$~MeV. In all of the cases the $N_t=6$ results were found
to be outside of the scaling region and have thus been excluded from the fit.
One of these extrapolations is visualized in Fig.~\ref{fig:tconti} for $\Delta^{\rm NLO}/m_\pi^3=0.61$. Finally, the
contour lines of $\Delta^{\rm NLO}$ in the continuum limit are plotted in Fig.~\ref{fig:tcontour-conti},
this time versus $\mu_I/T$. Once more, the gray area indicates the BEC phase in the continuum with the
updated phase boundary from App.~\ref{app:pcbound}. As discussed above, the naive expectation for the
contours outside of the BEC phase are lines with $\mu/T=\textmd{const}.$, i.e.\ vertical lines in this plot.
While this is approximately the case for large $\Delta^{\rm NLO}$, the contours with
small values of $\Delta^{\rm NLO}$ show
clear deviations from this expectation with the tendency to shift to larger values of $\mu_I/T$ with
increasing temperature.

\end{counted}\wc

\section{Testing the radius of convergence}
\label{sec:rconv}

\begin{counted}

As mentioned in the introduction, the radius of convergence of the Taylor expansion 
has been used extensively in the literature to extract information on the possible
phase transitions of the theory for $\mu_B>0$.
The current setup with $\mu_I>0$ is ideal to test the
performance of this method in QCD, since the phase diagram features a second order
phase transition~\cite{Son:2000xc,Brandt:2017oyy} comparably close to the $\mu_I=0$ axis. We have already
seen the breakdown of the expansion close to the phase boundary (cf. Fig.~\ref{fig:nt6-temps}).
We will now test whether the leading estimator for the radius of convergence also indicates the presence of this phase
boundary.

A possible definition for the radius of convergence $r$ for
the Taylor series of $\expv{n_I}$ from Eq.~(\ref{eq:nI-taylor}) is given by
\be
\label{eq:conv-rad-nI}
r = \lim_{n\to\infty} r_n(n_I), \quad\frac{r_n(n_I)}{T} = \sqrt{ \frac{c_{n}}{c_{n+2}} (n+1)n } \,,
\ee
but note that for a general singularity in the complex $\mu_I$-plane
this limit is not guaranteed to exist
(see Ref.~\cite{Vovchenko:2017gkg} for a counter-example).
Here $c_n$ are the coefficients of the expansion of the pressure defined in Sec.~\ref{sec:setup}.
Note that while the same radius of convergence $r$ is encoded in the Taylor series 
of other observables, the estimators at finite $n$ can be quite different. 
In particular, comparing the series for the pressure $p$, the density $\expv{n_I}$ 
and the susceptibility $\expv{\chi_I}=\partial \expv{n_I}/\partial\mu_I$ gives
\be
\label{eq:conv-rad-rel}
r_n(\chi_I)=\sqrt{\frac{n-1}{n+1}} \, r_n(n_I) = \sqrt{\frac{n(n-1)}{(n+2)(n+1)}} \, r_n(p) \,.
\ee
These indeed agree in the limit $n\to\infty$, but differ at finite $n$.

\end{counted}\wc

\begin{figure}[t]
 \centering
 \includegraphics[width=8cm]{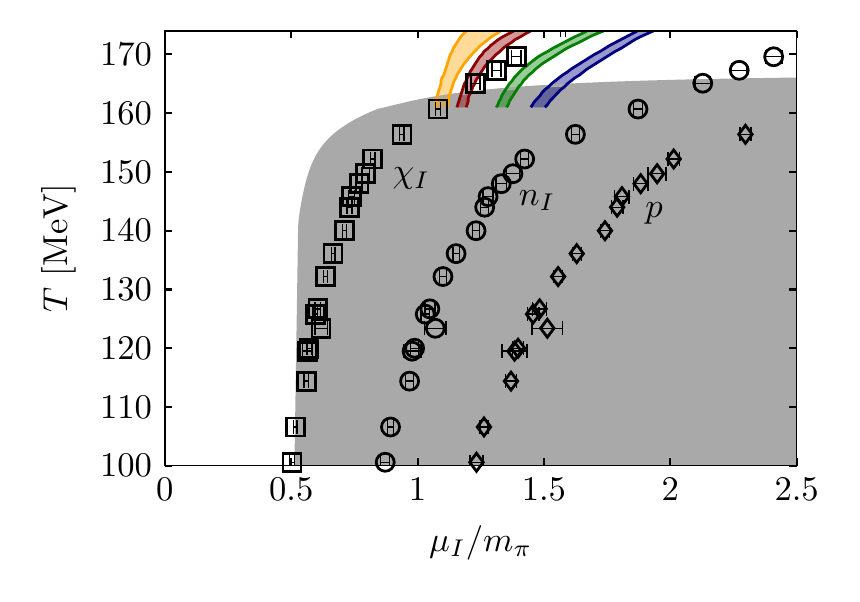}
 \caption{\label{fig:convrad}
 The leading-order estimators for the radius of convergence using the series for 
 different observables on our $N_t=8$ ensembles. The results are compared to the
 boundary of the BEC phase (gray area) 
 and to the contours of $\Delta^{\rm NLO}$ (colored bands).
 }
\end{figure}

\begin{counted}

With two coefficients at hand, only a single estimator can be constructed for $r$
and we cannot investigate the $n\to\infty$ limit systematically. The estimators for $r$ 
obtained from the different observables are
shown in Fig.~\ref{fig:convrad} for $N_t=8$. For comparison we also included the 
boundary of the BEC phase and the contour lines
of $\Delta^{\rm NLO}$ in the figure. Both the estimators $r_n$ and the $n$-th order 
contour lines are expected to fall on top of the phase boundary
in the limit $n\to\infty$ -- as long as the BEC onset is the singularity closest to $\mu_I=0$.

The results for $r_2(\chi_I)$ are observed to lie surprisingly close to 
the phase boundary for low temperatures, while $r_2(n_I)$ and $r_2(p)$ 
significantly overestimate the radius of convergence. 
While such a perfect agreement for $r_2(\chi_I)$ is likely accidental, 
similar tendencies were also found in an NJL-type model~\cite{Karsch:2010hm}, 
and in toy models of QCD with imaginary chemical potentials~\cite{DElia:2016jqh}, 
suggesting that the series of the susceptibility gives estimators with the 
fastest convergence rate.
The estimators also reveal a considerable change of slope around $150-160\textmd{ MeV}$, 
close to the upper boundary of the BEC phase and in agreement with the qualitative 
trend that the two curves will agree in the limit $n\to\infty$. Finally, 
a qualitative agreement is also observed 
between the behavior of $r_2$ and the contours lines of $\Delta^{\rm NLO}$.

Up to now we have ignored a subtle issue regarding
the estimators of the radius of convergence in finite volumes.
In a finite volume $V$, phase transitions are smoothed out,
but the partition function has
Lee-Yang zeroes at complex $\mu_I$
that approach the real axis as $V\to\infty$~\cite{Yang:1952be},
according to criticality~\cite{Stephanov:2006dn}.
How this is connected to the $V$-dependence of the estimators $r_n$
is highly non-trivial.
We find that the leading order  Taylor
coefficients depend only mildly on the volume. This is discussed in App.~\ref{app:ctaylor}
together with the finite size effects of the full data.

\end{counted}\wc

\section{Conclusions}
\label{sec:conc}

\begin{counted}

In this letter we have presented a detailed comparison between Taylor expansion and full simulations
of QCD at nonzero isospin chemical potential $\mu_I$. 
This is a theory with a second-order phase transition between the normal phase
and a phase with Bose-Einstein condensation of charged pions,
enabling us to observe the breakdown of the Taylor expansion at the critical chemical potential.
Up to the boundary of the BEC phase the full data for the isospin density $\expv{n_I}$
is well described by Taylor expansion -- both by the leading- (LO) and the next-to-leading order (NLO)
series.

To test the reliability of the expansion outside the BEC phase, we extended our 
lattice ensembles generated in~\cite{Brandt:2017oyy} with simulations at higher 
chemical potentials. The update for the BEC phase boundary up to $\mu_I\approx325$~MeV
is presented in App.~\ref{app:pcbound}. To quantify the performance of the LO and NLO expansions
in this region, we introduced the deviation $\Delta$ between the full and the 
Taylor-expanded results, see Eq.~(\ref{eq:taylor-dif}). The contour lines of 
$\Delta^{\rm LO}$ and $\Delta^{\rm NLO}$ are shown in Fig.~\ref{fig:tcontour-nt8}
for our $N_t=8$ ensembles. Taking the $\Delta=0.2\,m_\pi^3$ contour (which 
corresponds to deviations of $3-6\%$ in the isospin density) as an indicator, the LO expansion 
is reliable up to $\mu_I/m_\pi\approx 0.5$ to $0.6$, while the NLO series performs 
reasonably well up to $\mu_I/m_\pi\approx 1.0$ to $1.5$.
The continuum extrapolation of the contour lines is visualized in Fig.~\ref{fig:tcontour-conti}. Contrary to the 
naive expectation that the contours lie along lines of constant $\mu/T$ outside of the BEC phase,
we find considerable deviations from this behavior, with a tendency towards larger values
of $\mu/T$ with increasing temperature.

We have also compared the estimator $r_2$ for the radius of convergence, obtained from the first two coefficients
($c_2$ and $c_4$) of the Taylor expansion, to the critical chemical potential $\mu_{I,c}$
known from the full simulations. We find that both $r_2$ and $\mu_{I,c}$
change similarly with temperature, signaling the expected agreement when higher order estimates for
the radius of convergence are taken into account.
Concerning possible different definitions of $r$, we find that the estimate obtained from
the susceptibility $\chi_I$ is closest to the phase boundary. Whether this remains true for higher-order estimators
remains to be seen. Our findings demonstrate that already the leading estimators for 
the radius of convergence are sensitive to the phase transition. We emphasize,
however, that a more detailed study including higher order estimates for $r$ is clearly desired and mandatory
to be able to draw definite conclusions.
Our study can be used to quantify the uncertainties of Taylor expansions in 
baryon chemical potentials and to guide the interpretation of results obtained 
for the radius of convergence of such series.
This will also be of relevance for comparison to low energy models of QCD.\\

\end{counted}\wc

\noindent
{\bf Acknowledgments}
The authors thank Szabolcs Bors\'anyi for useful correspondence and for providing the data
for the Taylor expansion coefficients, and Volodymyr Vovchenko for insightful comments.
This research was funded by the DFG (Emmy Noether Programme EN 1064/2-1 and
SFB/TRR 55). The simulations were performed on the GPU cluster of the Institute for
Theoretical Physics at the University of Regensburg and on the FUCHS and LOEWE clusters
at the Center for Scientific Computing of the Goethe University of Frankfurt.

\appendixsection

\section{\boldmath Computation and improved $\lambda$-extrapolations of $n_I$}
\label{app:nI}

In terms of the massless Dirac operator $\Dp$ and the mass of the (degenerate) light quarks
$m_\light$, the isospin density~(\ref{eq:nI-def}) is given by
(cf. Ref.~\cite{Brandt:2016zdy})
\be
\label{eq:nI-op}
\expv{n_I} = \frac{T}{2V} \expv{ \textmd{Re }\tr \frac{(\Dp+m_\light)^\dagger
\partial_{\mu_I}\Dp}{|\Dp+m_\light|^2+\lambda^2}} \,,
\ee
where $\partial_{\mu_I}$ is the derivative with respect to $\mu_I$. The trace in
Eq.~(\ref{eq:nI-op}) can be evaluated using stochastic estimators, giving $n_I^{\rm stoch}$, or in the spectral representation
\be
\label{eq:nI-spect}
n_I = \frac{T}{2V} \sum_n \frac{\textmd{Re}\; \varphi_n^\dagger [\Dp+m_\light]^\dagger \partial_{\mu_I}\Dp
\varphi_n}{\xi^2_n+\lambda^2}\,,
\ee
with the singular values $\xi_n$ and the associated eigenstates $\varphi_n$
(see Ref.~\cite{Brandt:2017oyy} for more details).

The spectral representation~(\ref{eq:nI-spect}) is the basis for the improvement
program introduced in Ref.~\cite{Brandt:2017oyy}. First we introduce the truncated difference
\be
\label{eq:ldiffs-trunc}
\begin{split}
\dNn &\equiv n_I^{\Nxi}(\lambda)-n_I^{\Nxi}(\lambda=0) \\
& 
\begin{split}
= \frac{T}{2V} \sum_{n=1}^{\Nxi} & \textmd{Re}\; \varphi_n^\dagger [\Dp+m_\light]^\dagger \partial_{\mu_I}\Dp
\varphi_n \\
&\times \left( \frac{1}{\xi_n^2+\lambda^2}-\frac{1}{\xi_n^2}\right)\,,
\end{split}
\end{split}
\ee
where $n_I^{\Nxi}(\lambda)$ is the operator from Eq.~(\ref{eq:nI-spect}) with a singular value sum truncated
at $n=\Nxi$.
This truncated difference $\dNn$ does not contribute in the $\lambda\to0$ limit, allowing 
us to write
\be
\label{eq:liml0}
\lim_{\lambda\to0} \expv{n_I}=\lim_{\lambda\to0} \expv{n_I^{\rm stoch}-\dNn} \,.
\ee
As indicated, the $\lambda>0$ value of the operator is determined using 
stochastic estimators, while the correction term $\delta^N$ is calculated 
in the spectral representation~(\ref{eq:ldiffs-trunc}). 
In addition to the improvement of the operator, we also employ the leading order 
reweighting discussed in section III.4 of
Ref.~\cite{Brandt:2017oyy}. This approximates the $\lambda=0$ 
distribution of the lattice ensembles and brings the expectation value $\expv{n_I}$ 
closer to its $\lambda=0$ limit. 

\begin{figure}[t]
 \centering
 \includegraphics[width=7.6cm]{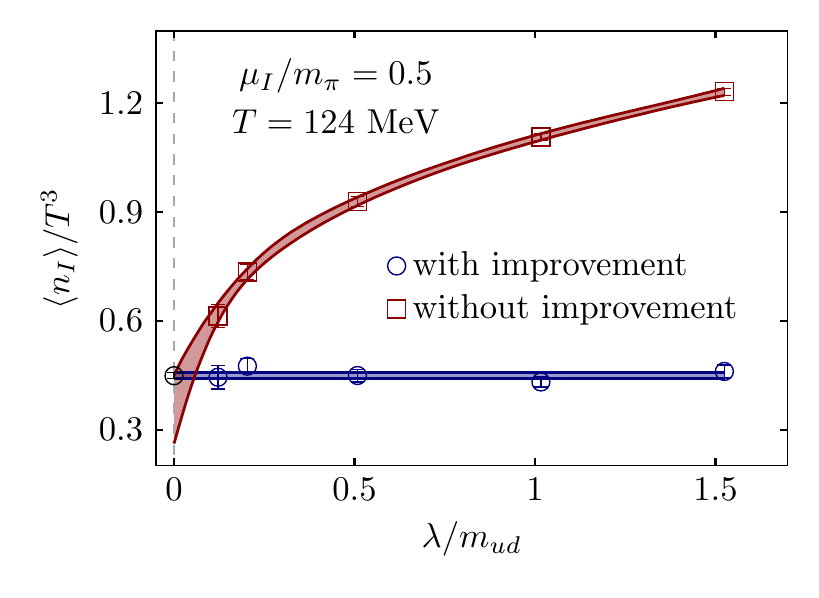}
 \caption{\label{fig:lextra}
 Improved $\lambda$-extrapolation for $\expv{n_I}$ on our $24^3\times6$ ensemble in comparison
 to the unimproved one.
 }
\end{figure}

As discussed in Ref.~\cite{Brandt:2017oyy}, the optimal (or minimal) value of $N$ necessary to achieve sufficient
improvement to obtain a controlled $\lambda$-extrapolation will in general depend on the operator. We find
that the behavior of $n_I$ with $N$ is similar to the one of the chiral condensate, so that $N\approx100$
is usually sufficient to obtain reasonably flat extrapolations.
A particular example for the $\lambda$-extrapolations is provided in Fig.~\ref{fig:lextra}. The plot indicates
the tremendous increase in reliability due to our improvement scheme, enabling a well controlled
$\lambda$-extrapolation and precision results.

\section{Interpolation of Taylor expansion coefficients and finite size effects}
\label{app:ctaylor}

\begin{figure}[t]
 \centering
 \vspace*{-.2cm}
 \hspace*{.7cm}\includegraphics[width=7.2cm]{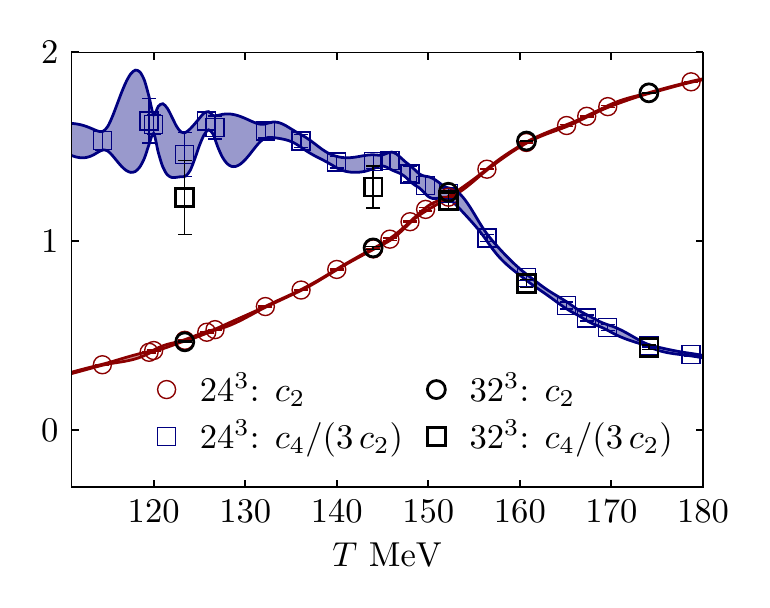}\quad
 \includegraphics[width=8cm]{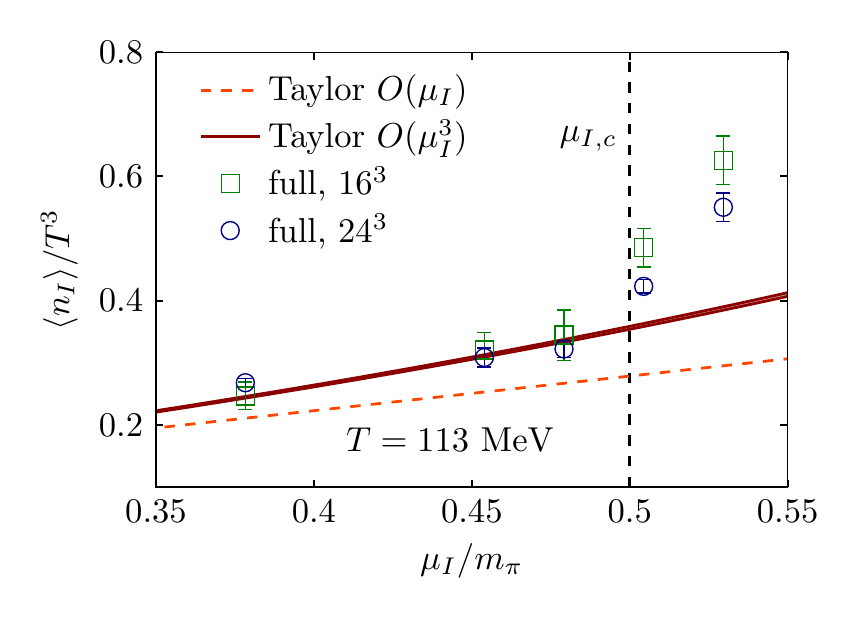}
 \caption{\label{fig:tcoeff}
 Top panel:
 Spline interpolation of the first two Taylor coefficients $c_2$ and $c_4$ (normalized by
 $3c_2$) on the $24^3\times8$ lattices,
 compared to the results on the $32^3\times8$ ensemble~\cite{Borsanyi:2011sw}.
 Bottom panel:
 Results for $\expv{n_I}$ on $N_t=6$ lattices from direct simulations with a spatial
 volume of $16^3$ (green boxes) and $24^3$ (blue circles)
 in comparison to LO (orange dashed line) and NLO (red solid line) Taylor expansions
 for $T=113 \textmd{ MeV}$. The Taylor
 expansion coefficients have been obtained on a $18^3\times6$ lattice~\cite{Borsanyi:2011sw}.
 }
\end{figure}

The Taylor coefficients are combinations of derivatives of the pressure,
\be
\label{eq:p-def}
c_n = \left.\frac{\partial^n (p/T^4)}{\partial (\mu_I/T)^n}\right|_{\mu_I=0}\,, \quad\quad
\frac{p}{T^4} = \frac{1}{VT^3} \log \Z \,,
\ee
with respect to the isospin chemical potential at $\mu_I=0$.
These can be rewritten using the quark chemical potentials $\mu_u$ and $\mu_d$ --
in particular, $c_2$ and $c_4$ are given by
\be
\label{eq:taylor-c2}
c_2=2 \left. \Big[\partial_u^2\Big(\frac{p}{T^4}\Big) - \partial_u \partial_d
\Big(\frac{p}{T^4}\Big) \Big] \right|_{\mu_I=0}
\ee
and
\be
\label{eq:taylor-c4}
c_4=2 \left. \Big[ \partial_u^4
\Big(\frac{p}{T^4}\Big) - 4 \partial_u^3 \partial_d \Big(\frac{p}{T^4}\Big) + 3
\partial_u^2 \partial_d^2 \Big(\frac{p}{T^4}\Big) \Big] \right|_{\mu_I=0} \,,
\ee
where $\partial_f$ stands for the derivative with respect to $\mu_f/T$. The results for the coefficients
$c_2$ and $c_4$ from Ref.~\cite{Borsanyi:2011sw} for the $24^3\times8$ lattice are shown in the top panel
of Fig.~\ref{fig:tcoeff} together with a cubic spline interpolation.

To check finite size effects in the temperature range of interest we 
also include the results from the $32^3\times8$ ensemble from Ref.~\cite{Borsanyi:2011sw} in
the top panel of Fig.~\ref{fig:tcoeff}.
Apart from some visible but not significant effects for the $c_4$ coefficient at $T\lesssim150$~MeV
finite size effects are absent. To lend further support to the statement that finite size effects are
negligible, we show the results for $\expv{n_I}$ obtained on $16^3\times 6$ and $24^3\times6$ lattices
in comparison to the Taylor expansion with coefficients from a $18^3\times6$ lattice in the 
bottom panel of Fig.~\ref{fig:tcoeff}. 
For $\mu_I<\mu_{I,c}$, we see that the lattice results agree within uncertainties and are in mutual agreement
with the NLO expansion.
Also evident are the expected strong finite size effects at and just above $\mu_{I,c}$, outside 
the applicability region of the expansion.

The main part of our study has been done on $24^3\times6$, $24^3\times8$, $28^3\times10$ and $36^3\times12$ lattices. In
contrast the Taylor coefficients have been computed on $18^3\times6$, $24^3\times8$,
$32^3\times8$, $32^3\times10$ and $32^3\times12$ lattices in Ref.~\cite{Borsanyi:2011sw}. Given the magnitude
of finite size effects visible in Fig.~\ref{fig:tcoeff}, we can thus conclude that those effects are irrelevant
within the present accuracy.

\section{\boldmath The pion condensation phase boundary for large $\mu_I$}
\label{app:pcbound}

\begin{figure}[t]
 \centering
 \includegraphics[width=8cm]{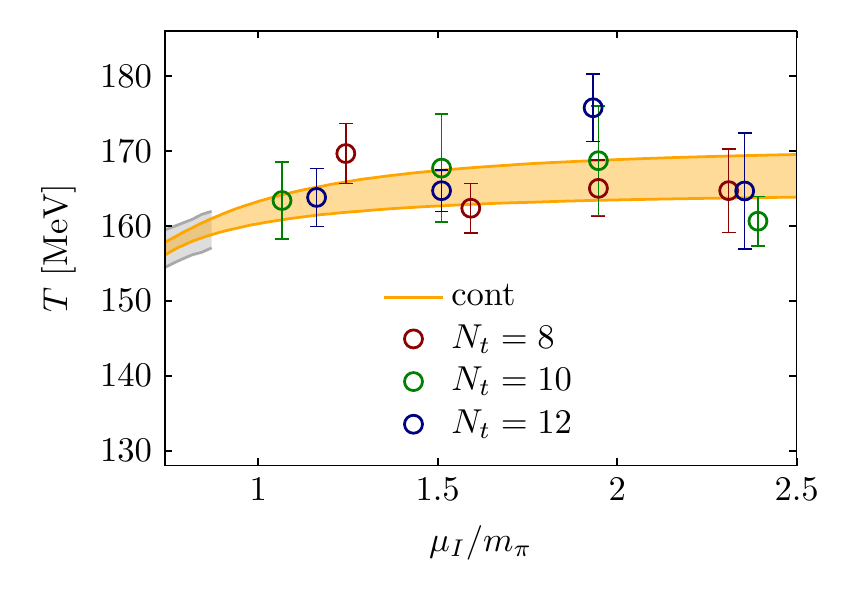}
 \caption{\label{fig:bec-conti}
 Continuum extrapolation (yellow band) of the BEC phase boundary based on results from three lattice 
 ensembles (colored points).
 The gray band is the part of the continuum extrapolation from Ref.~\cite{Brandt:2017oyy} which
 enters the fit for the purpose of matching to the phase boundary for $\mu_I<120$~MeV.
 }
\end{figure}

For the present study we extended the range of chemical potentials compared to 
Ref.~\cite{Brandt:2017oyy}, enabling a determination of the BEC phase boundary for higher values
of $\mu_I$. We have performed new temperature scans in the range $120\textmd{ MeV}<\mu_I< 325$~MeV, 
allowing to locate the critical temperature $T_c(\mu_I)$, where the pion condensate 
vanishes. The results for three lattice spacings are shown in Fig.~\ref{fig:bec-conti}.
Compared to our previous results~\cite{Brandt:2017oyy} we observe a slight increase
in the critical temperature with all data points approximately lying along a constant line. 
To capture this behavior and to approach the continuum limit, we fit all points by the function
$d_1 + d_2/\mu_I^2$ with $a^2$-dependent coefficients $d_1$ and $d_2$. To smoothly connect to our previous
result, the data of~\cite{Brandt:2017oyy} for $T<161\textmd{ MeV}$ and 90~MeV
$\leq \mu_I \leq 120$~MeV for each value of $N_t$ and in the continuum are also included in the fit.
The resulting continuum extrapolation is also shown in Fig.~\ref{fig:bec-conti}.

\providecommand{\href}[2]{#2}\begingroup\raggedright\endgroup

\end{document}